# STABILIZATION OF UNSTABLE PERIODIC ORBITS IN DC DRIVES

KRISHNENDU CHAKRABARTY[1]

*Kalyani Government Engineering College*
*Kalyani-741235, India[2]*
chakrabarty40@rediffmail.com

URMILA KAR[3]

*National Institute of Technical Teachers Training and Research,*
*Kolkata-700106, India[1]*
urmilakar@rediffmail.com

Electric drive using dc shunt motor or permanent magnet dc (PMDC) motor as prime mover exhibits bifurcation and chaos. The characteristics of dc shunt and PMDC motors are linear in nature. These motors are controlled by pulse width modulation (PWM) technique with the help of semiconductor switches. These switches are nonlinear element that introduces nonlinear characteristics in the drive. Any nonlinear system can exhibit bifurcation and chaos. dc shunt or PMDC drives show normal behavior with certain range of parameter values. It is also observed that these drive show chaos for significantly large ranges of parameter values. In this paper we present a method for controlling chaos applicable to dc shunt and PMDC drives. The results of numerical investigation are presented.

*Keywords*-Chaos, Control of chaos, DC drive.

## 1. Introduction

There is a myth that all oscillations are periodic in nature. But nonlinear system shows oscillatory behavior under certain parameter values that are not periodic but bounded in the phase plane. This behavior is inherent to a nonlinear system. This aperiodic behavior is not due to any noise or due to any external interference. This behavior is called chaotic. Hence chaos is aperiodic long-term behavior of a system having definite mathematical formulation. The positive lyapunov exponents of the chaotic system quantify the existence of chaos. PMDC motors are useful in a range of applications. To name a few, it is used in battery powered device like wheel chair, power tools, conveyors, door openers and pumping equipments.

K.T Chau et al have studied the bifurcation and chaos in voltage controlled PMDC drives. [1]. Similar study with current controlled PMDC motor has been performed by J.H Chen el al with simulation and experimental validation.

DC series motor also has very wide applications in electric traction. The detailed study of dynamics of DC series motor has been reported. in [3,4].

The strange attractor of a chaotic system is a collection of infinite periodic orbits embedded in the system's the phase plane. All these periodic orbits are in the unstable state. As for example, the trajectory in the chaotic attractor comes very near to period-1 orbit and goes away with time without becoming stable period-1 orbit. The same is observed for other period-P orbits. This characteristic of chaotic system makes it possible to stabilize any unstable periodic orbit with application of tiny change of parameters or by introduction of control algorithm. In recent years, investigators have shown that by perturbing a chaotic system in right way, one can force the system to follow one of its many unstable behaviors. With proper control one can rapidly switch among many different behaviors. This gives a clue to improve the response as well as the domain of operation in systems that exhibit chaos for some parameter values.

Various methods to control chaos in different chaotic systems have been developed. An overview of different approaches to the control of chaos for various nonlinear dynamic systems has been reported by G.Chen el al [5].

The PWM controlled dc shunt drives under investigation is linear in nature but bifurcation and chaos occurs in this system due to switching nonlinearity. Hence there was a need to develop a method suitable for circuits with switching nonlinearity. Methods for typical systems with switching nonlinearity have been proposed in [6, 7, and 8]. These methods take the advantage of the existence of unstable periodic orbits in the strange attractor as in the OGY method. But this method differs from the OGY method in the control logic.

J.H. Chen et al used delayed self-controlling feedback to stabilize PMDC drive [9]. This enables to stabilize the dynamics of the system to period-1 or any sub harmonic period.

In this work, an attempt is made to develop a simple control algorithm, using linear structure of the dc shunt drive before and after switching to control any unstable periodic orbit of the chaotic attractor. The control strategy is to find the fixed point corresponding to the periodic orbit that to be stabilized. Once the trajectory comes very near to the target fixed point, control is applied in the form of a slight change of any parameter so that the chaotic trajectory follows the target trajectory and stabilizes to that.

2. **The system.**

The block diagram and equivalent circuit of the dc shunt drive circuit is shown in the Fig.1 and Fig.2 respectively. The output of the speed sensor (W) is compared with the reference speed ($W_{ref}$) in the comparator $A_1$. The difference of (W- $W_{ref}$) is compared with a ramp voltage. The output of the comparator $A_2$ is used to switch on the switch for the chopper drive. The system is linear in nature except the switch. The chaotic behavior of the system is due to switching nonlinearity. The SCR in the circuit is used as a switch.. The commutation circuit of the SCR is not shown in the figure. The switch can be implemented by any other semiconductor switch e.g. MOSFET etc.

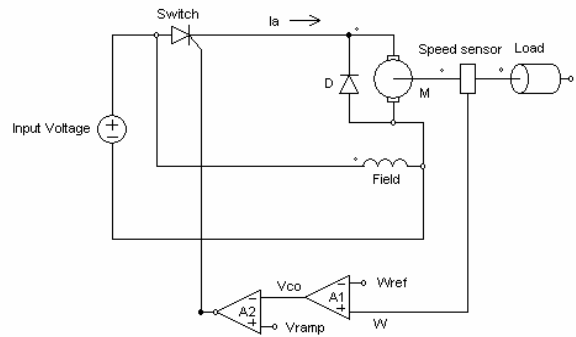

Fig 1: Block diagram of dc shunt drive

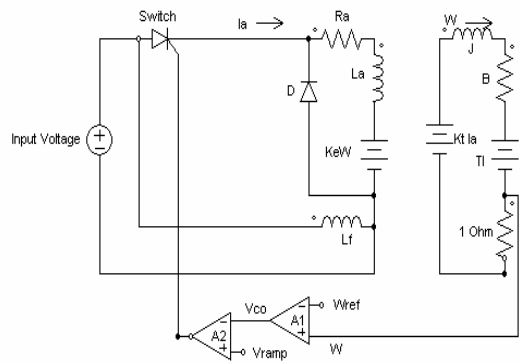

Fig 2: Equivalent circuit of dc shunt drive

## 3. Modeling of the system

As shown in Fig.1 a simple dc chopper fed dc motor drive operating in the continuous current mode is used. This corresponding equivalent circuit as is shown in Fig.2 will be used throughput for analysis.
The armature current and speed of the dc shunt drive have been taken as state variables. The system can be divided into two stages depending on the switching conditions.

The switch will be on when $V_{co}<V_{ramp}$. The corresponding system equations are:

$$\frac{d}{dt}\begin{pmatrix} i(t) \\ \omega(t) \end{pmatrix} = \begin{pmatrix} \frac{-R}{L} & \frac{-K_E}{L} \\ \frac{K_T}{J} & \frac{-B}{J} \end{pmatrix} \begin{pmatrix} i(t) \\ \omega(t) \end{pmatrix} + \begin{pmatrix} \frac{V}{L} \\ \frac{-T_l}{J} \end{pmatrix}$$

The switch will be off when $V_{co}>V_{ramp}$. The corresponding equations are:

$$\frac{d}{dt}\begin{pmatrix} i(t) \\ \omega(t) \end{pmatrix} = \begin{pmatrix} \frac{-R}{L} & \frac{-K_E}{L} \\ \frac{K_T}{J} & \frac{-B}{J} \end{pmatrix} \begin{pmatrix} i(t) \\ \omega(t) \end{pmatrix} + \begin{pmatrix} 0 \\ \frac{-T_l}{J} \end{pmatrix}$$

Where, i (t) is armature current. R armature resistance, L armature inductance, V dc supply voltage, $K_E$ back emf constant, $K_T$ torque constant, B viscous damping, J load inertia and $T_l$ load torque.

By defining the state vector X (t) and the following matrices A, $E_1$, $E_2$

$$A = \begin{pmatrix} \frac{-R}{L} & \frac{-K_E}{L} \\ \frac{K_T}{J} & \frac{-B}{J} \end{pmatrix}, X(t) = \begin{pmatrix} i(t) \\ \omega(t) \end{pmatrix}$$

$$E_1 = \begin{pmatrix} \frac{V}{L} \\ \frac{-T_l}{J} \end{pmatrix}, E_2 = \begin{pmatrix} 0 \\ \frac{-T_l}{J} \end{pmatrix}$$

The system can be rewritten as
$$\dot{X}(t) = AX(t) + E_K \quad ----(3)$$
Where K= 1, 2. K changes the value depending on ON or OFF condition of the switch.
Thus, the closed loop drive system is a second order non-autonomous dynamical system.
Given the desired initial condition X ($t_0$), the analytical solution of the system equation given by (3) can be expressed as

$$X(t) = \phi(t-t_0)X(t_0) + \int_{t_0}^{t} \phi(t-\tau)E_K d\tau$$

$$= -A^{-1}E_K + \phi(t-t_0)(X(t_0) + A^{-1}E_K) \quad ----(4)$$
Where K=1, 2 and $\varphi(t) = e^{At}$ is state transition matrix.

## 4. The method of stabilization of chaos

Let $X_{on}(1)$ and $X_{on}(2)$ be the values of armature current and speed respectively at the instant of switching, $X_{off}(1)$ and $X_{off}(2)$ are the corresponding values `at switch off. We define

$$XON = \begin{pmatrix} X_{on}(1) \\ X_{on}(2) \end{pmatrix}, XOFF = \begin{pmatrix} X_{off}(1) \\ X_{off}(2) \end{pmatrix}$$

The dynamics of the system can be expressed as
$$XON_{n+1} = f(XON_n, P)$$
where P is accessible parameter that can be varied to achieve stabilization.
The parameter P needs to be changed at the instant when the trajectory comes very near to fixed point corresponding to the unstable periodic orbit that need to be stabilized.
If period-1 is to be stabilized, then period-1 fixed point at switch on and off can be found out from the condition
$$XON_{n+1} = XON_n, XOFF_{n+1} = XOFF_n$$
which is denoted by $XON^*$ and $XOFF^*$. The time required to travel from XON to (in the neighbor hood of $XON^*$) $XOFF^*$.is known from the equation of ramp wave. Thus it is a problem of targeting $XOFF^*$ from XON by change of the parameter P determined with the help of equation (4).

## 5. Result

The method of control of chaos is verified numerically with the following parameters:
B=0.000275Nm/rads$^{-1}$, J=0.00557Nm/rads$^{-2}$, L=53.7mH, $K_T$=0.1324Nm/A, $K_E$=0.1356Nm/rads$^{-2}$, R=2.8 ohm, $W_{ref}$=100 rads$^{-1}$, $T_l$=0.38 Nm, The ramp varies from 0 to 2.2 volts in 10 ms.
The method of stabilization of chaos developed is general in nature. It can stabilize the period-1 and other sub harmonic trajectory with period-P. In this work we have only stabilized period-1 and period-2 orbits.
The Fig. 3 Shows the time plot of armature current and speed at chaotic condition with input voltage 151 V. The corresponding space plot is shown in Fig. 4. When the trajectory comes in the close neighborhood of fixed point, control is applied in the form of slight change of input voltage to the system. as shown in figure 5, 6, & 9 The input voltage is taken as control parameter. The stabilized Period-2 orbit showing armature current and speed in the time axis along with phase plot is shown in Fig. 5,6 & 7 respectively. The DC shunt drive also shows chaotic behavior at 152 V. Fig. 8 show the chaotic attractor at input voltage 152.V. The stabilized period-1 time plot of armature current is shown in Fig.9. The corresponding phase plot is shown in Fig. 10.

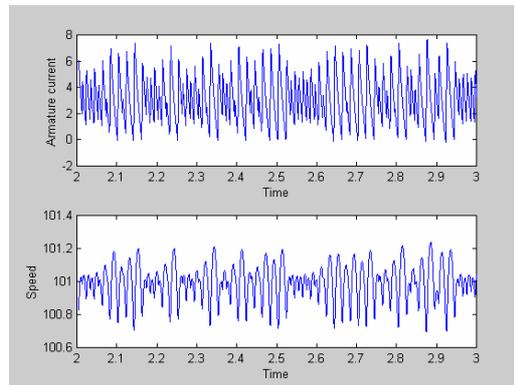

Fig 3: The time plot of current, speed under chaotic condition with input voltage of 151 V.

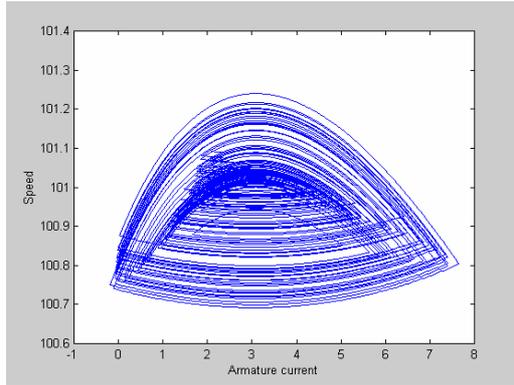

Fig 4: The phase plot of current and speed with input voltage of 151 V under chaotic condition.

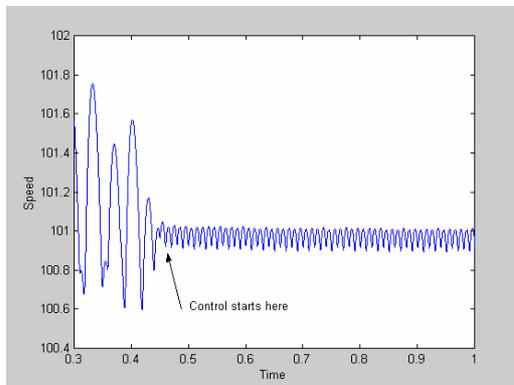

Fig 5: Time plot of speed after control of period-2 for input voltage of 151 V.

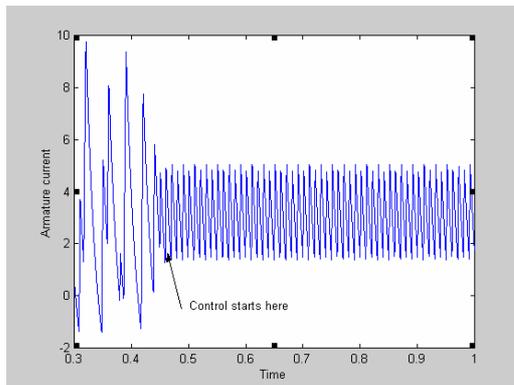

Fig 6: Time plot of armature current after control of period-2 for input voltage of 151 V.

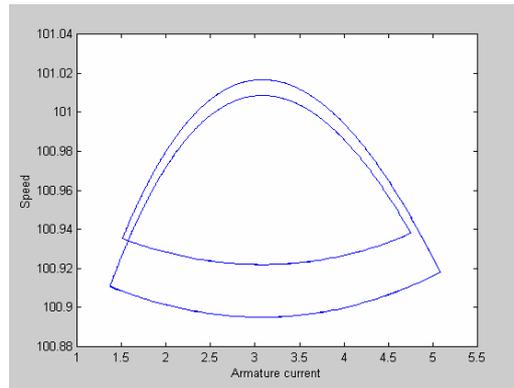

Fig 7: Phase plot of Armature Current and speed after control of period-2 for input voltage of 151 V.

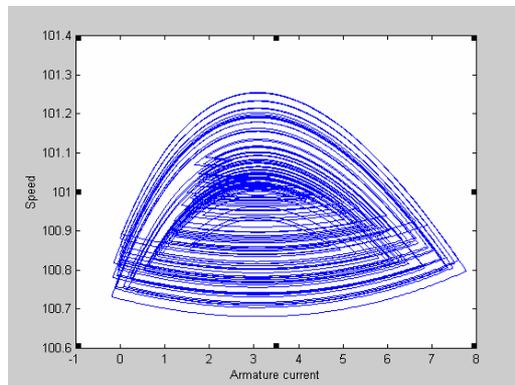

Fig 8: Phase plot of armature current and speed at input voltage of 152 V under chaotic condition

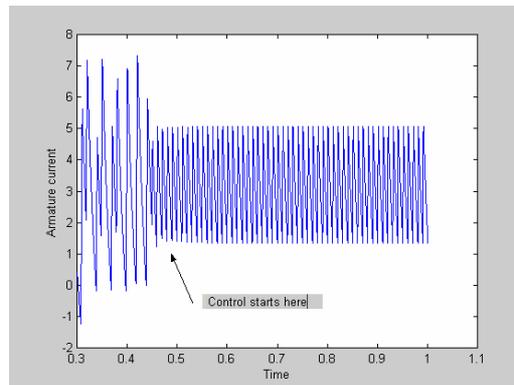

Fig 9: Time plot of armature current at input voltage of 152 V after control of Period-1.

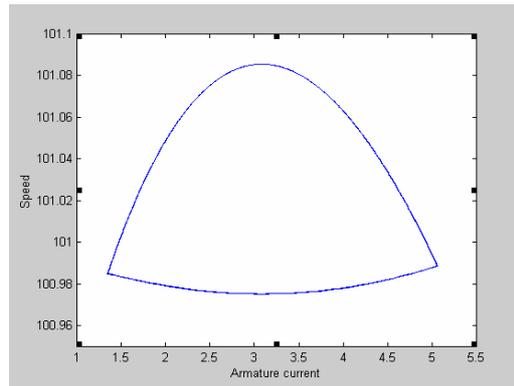

Fig. 10: Phase plot of armature current Vs speed at input voltage of 152 V after control of period-1.

6. **Conclusion**

It has been shown that the unstable periodic obit of chaotic attractor of dc drives can be stabilized by targeting the fixed points corresponding to the unstable periodic orbit that need to be stabilized with the help of slight variation of a desired parameter of the system .. This method of control of chaos is suitable to systems like dc drives that are linear except during switching nonlinearity. The computation of the appropriate value of the parameter that needs to be applied to stabilize a particular orbit is simple. It requires a state equation to be solved.

---

i